\documentclass[preprint,twoside,dvips]{aastex}
\usepackage{graphicx}

\begin{document}        
\title{KINEMATICS OF 1200 KM/S JETS IN HE 3-1475}
\author{Kazimierz J. Borkowski}
\affil{Department of Physics, North Carolina State         
University, Raleigh, NC 27695}
\email{kborkow@unity.ncsu.edu}        
\and
\author{J.~Patrick Harrington}
\affil{Department of Astronomy, University of Maryland, College Park, MD 20742}
\email{jph@astro.umd.edu}

\begin{abstract}
Spectroscopic observations of a proto-planetary nebula He 3-1475 with 
the Space Telescope Imaging Spectrograph (STIS) reveal the kinematics of
its high (1200 km s$^{-1}$) velocity jets. The jets are formed at a large 
(0.15 pc) distance from its central star by collimation of an asymmetric
stellar wind in a pair of conical shocks seen in Wide Field Planetary Camera 
(WFPC2)
images. The jets consist of several pairs of knots symmetrically distributed
with respect to the central star, with most knots exhibiting a head-tail
morphology. Large (up to 650 km s$^{-1}$) radial velocity
gradients are seen within the knots on subarcsec scales, with velocities
decreasing from the knot heads toward their trailing tails. These large velocity
gradients are a sign of efficient deceleration of jets by a much slower bipolar
outflow. The inclination angle of the bipolar outflow is equal to 40\degr,
based on Doppler shifts of the scattered stellar H$\alpha$ line. Its velocity
is equal to 140 km s$^{-1}$ at a distance of 0.23 pc from the star, and 
increases monotonically with the radial distance from the star. A comparison 
of new WFPC2 [N II] $\lambda$6584 images 
with older WFPC2 images reveals expansion of the jets. The measured jet
proper motions in combination with their radial velocities imply
that He 3-1475 is a Galactic Bulge star at a distance of 8 kpc, located
800 pc above the Galactic plane. Its very high luminosity (25,000 $L_\odot$)
implies that He 3-1475 must be significantly more massive than a typical
AGB star within the Galactic Bulge, perhaps because of a past mass transfer 
and/or a merger event in an interacting binary system.

\end{abstract}        
        
\section{INTRODUCTION}

He 3-1475 is a B[e] type star located in the direction of the Galactic Bulge,
at $l = 9$\fdg4, $b = +5$\fdg8. Stars of this spectral type do not show 
typical absorption line spectra, but instead numerous permitted and forbidden 
emission lines are superposed on featureless continua (for further information
see a recent review on B[e] stars by Zickgraf 1998). 
He 3-1475 is an unusual B[e] star because
it is surrounded by a dense expanding torus of 
circumstellar matter (CSM) seen in OH maser
emission (Bobrowsky et al. 1995), and also as a prominent dark torus in Hubble
Space Telescope (HST) images (Bobrowsky et al. 1995; Borkowski et al. 1997).
A bipolar outflow and spectacular jets, perpendicular to the torus, were
detected in the optical by Riera et al. (1995) and Bobrowsky et al. (1995).
Radial velocities of bright knots in jets span 1000 km s$^{-1}$,
which is an order of magnitude more than what is typically
seen in both young and evolved stars surrounded by dense CSM. The morphology
of the jets is also unusual, because they do not originate in the central
star. Instead, a fast bipolar stellar outflow is apparently collimated 
far from the star in conical shocks seen in the HST images 
(Borkowski et al. 1997). The jets above the tips of conical shocks consist of 
a series of bright pairs of 
knots, located symmetrically with respect to the central star. Their spectra
show the presence of substantial ($\sim 450$ km s$^{-1}$) velocity gradients
within the knots, which indicates their strong deceleration by the ambient, 
more
slowly moving material in the bipolar lobes (Riera et al. 1995). There is
also evidence for deceleration of knots on large scales, as the knot radial
velocities decrease with increasing distance from the cone tips to 
the periphery of the bipolar outflow.

He 3-1475 is clearly an unusual source whose nature is not understood at
present. It's distance has been estimated at 1 -- 5 kpc (Riera et al. 1995;
Bobrowsky et al. 1995), which together with its high infrared flux
(Parthasarathy \&\ Pottasch 1989)
suggests that He 3-1475 is probably a post-Asymptotic Giant Branch (AGB) 
proto-planetary nebula (PPN). (Knapp et al. 1995 detected 8.4 GHz continuum radio 
emission, which they interpreted as coming from a compact, newly formed 
PN close to the central star.) This conclusion should be 
considered preliminary because a reliable distance determination is lacking.
The geometry of the circumstellar medium is also poorly known, with an
uncertain (30\degr\ -- Riera et al. 1995) inclination of the jets with 
respect to the line of sight. 
The velocity of the bipolar outflow and its age are unknown. This lack of 
fundamental knowledge about He 3-1475 makes it difficult to understand this
puzzling object. 

We report here on imaging and spectroscopic observations of He 3-1475 with
the HST, which allow us to determine its distance, inclination and
kinematics of its jets,
and the velocity of the bipolar outflow. Imaging and astrometry with the
Wide Field Planetary Camera (WFPC2) and the Space
Telescope Imaging Spectrograph (STIS) data are presented and discussed in 
\S~2 and \S~3, respectively. The jet inclination and the distance to He 3-1475
are determined in \S~4 and \S~5, and the nature of He 3-1475 and of its jets are
discussed in \S~6.

\section{WFPC2 IMAGING AND ASTROMETRY}

The HST observed He 3-1475 on September 9, 1999, in a number of narrow-band 
WFPC2 filters, with the nebula located entirely on the Planetary Camera
(PC) chip. We discuss here observations in the F658N filter
encompassing the [N II] $\lambda$6584 emission line, with the purpose of
determining proper motion of brightest knots in the jets. Two pairs of
images, shifted by 5.49 and 5.36 PC pixels along the horizontal and
vertical axes, with each image 
300 sec in duration, allows us to achieve nearly full HST spatial resolution.
After standard HST pipeline calibration performed at the Space Telescope 
Science Institute, we identified cosmic rays in each pair of images using
the ``crrej'' task from the Space Telescope Science Data Analysis System
(STSDAS) software. We 
then combined all 4 [N II] images using variable-pixel linear reconstruction
(``drizzling'') as implemented in the STSDAS ``dither'' package 
(Fruchter \& Hook 1998).
The final combined image (Figure 1a) has better spatial resolution and better
signal-to-noise (S/N) ratio than the previous 600 sec [N II] image. 
Three pairs of prominent knots, the inner, middle, and outer pairs 
labeled by (I, I'), (M, M'), and (O, O'), respectively, can be seen along 
the He 3-1475 jets. The SE jet at the top of Figure 1 is redshifted, while
the NW jet at the bottom is blueshifted.
Knots I and I' are located at or beyond the tips
of conical shocks where the collimation of jets presumably takes place. 

In order to determine the jet proper motion, we compared our new [N II] image 
with the previous HST [N II] image. This 1996 image consists of one pair of 
spatially coincident PC subimages, 400 sec and 200 sec in duration. 
Using STSDAS software, we found the rotation angle of 31.683 degrees between 
the 1999 and 1996 images, and then rotated and aligned the 1996 image 
with the 1999 image. 
The 1999-1996 [N II] difference image, corrected for the difference in 
exposure times, is shown in Figure 1b. 
This difference image clearly reveals the jet motion. Except for
the bright inner region, where the artifacts from the imperfect subtraction
of the central star dominate, all bright knots within the jet shifted outward
from the star, as shown by a characteristic difference pattern with negative 
values closer to the star and with positive values on the outside.
This radial motion is most
prominent for knot pairs (M, M') and (I, I'), although expansion is also
seen for the blue-shifted knot O'. (The red-shifted knot O is certainly 
also expanding, but this knot is affected by a bad PC column.) Motion of
fainter knots and of individual features within the knots is also seen.
For example, at least three features are seen moving
outward in the brightest knot (the blue-shifted knot I' in the lower half
of Figure 1), preceded on the outside by a bow-shock. Outward motion is also 
detected in the inner red-shifted jet, starting nearly at the base of the 
conical shock, and continuing along its knotty structure.

\begin{figure}    
\epsscale{0.95}    
\plotone{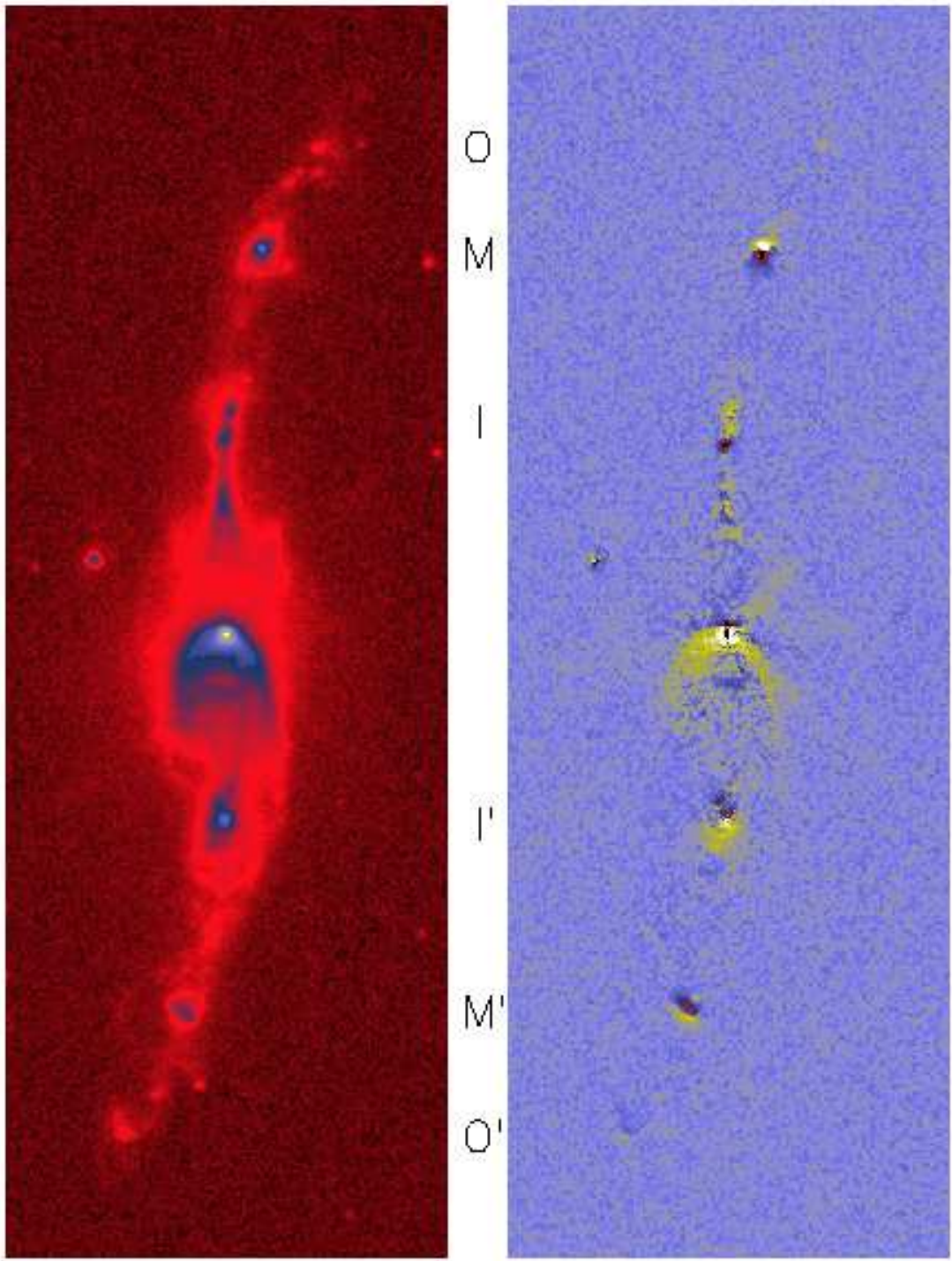}        
\caption{Protoplanetary nebula He 3-1475 imaged with the {\it Hubble Space 
Telescope} WFPC2 through the F658N ([N\,II] $\lambda$6584) filter: a) the
new 1200 sec image, b) the 1999-1996 difference image. Each image is
6\farcs8 $\times$ 19\farcs3 in size. The vertical axis is aligned with the
STIS slits oriented at position angle of 135\degr\ (SE is at the top, 
NW at the bottom). }
\label{fig::wfpc2}        
\end{figure}        

For quantitative measurements of proper motion, we chose the bright knot
pair (M, M'). These knots are clearly more suitable for this purpose than 
fainter knots (O, O'), and morphologies of these knots are considerably less 
complex that that of knots (I, I'). There is also less contamination from 
reflected 
stellar light at larger
distances from the central star, which scatters from dense, more slowly moving 
material in the bipolar lobes of He 3-1475. We used a cross correlation
technique to determine proper motions (Currie et al. 1996). Each pair of 
the 1999 images was
combined with the ``crrej'' task to form two cosmic-ray cleaned images to 
be compared with the cosmic-ray cleaned 1996 image. We then chose 
image subsections encompassing each of knot, and tapered their
edges in order to avoid any edge effects. With the rotation angle determined 
as explained previously, we cross correlated these subimages, using
a Fourier technique as implemented in the ``dither'' package. For the
central star, we found displacement between the images by fitting 
a two-dimensional Gaussian to the stellar image. The relative shifts between 
the knots and the central star were then averaged for the pair of 1999 images. 
We obtained radial
shifts of 0.73 and 0.81 PC pixels, and tangential (counterclockwise) shifts of 
0.00 and 0.12 pixels, for the red-shifted and blue-shifted knots M and M',
respectively. Statistical errors are just a few hundredths of a PC pixel,
but real errors dominated by systematic effects are likely to be several
times larger (Currie et al. 1996), maybe as large as 0.1--0.2 PC pixels.
We also attempted to determine the displacement for knots I and I',
arriving at radial displacements of 0.8 and 0.4 pixels for the brightest 
features
in the red-shifted and the blue-shifted knots I and I'. These inner knots 
displacements are very uncertain because of the complex morphology of  
knots I and I' and the presence of a more slowly moving material seen in the 
reflected
stellar light. Additional difficulties may arise for a highly 
blue-shifted knot I', where the [N II] emission falls on a rapidly changing
F658 filter profile.

The radial displacements of the red-shifted and blue-shifted knots M and M',
0.73 and 0.81 PC pixels (one PC pixel is equal to 45 miliarcsec), during the
time interval of 3.24 years between the observations, imply proper motions 
of 10 and 11 miliarcsec yr$^{-1}$. The kinematic ages of these knots can 
be obtained by dividing their distances from the central star, equal to
5\farcs95 for knot M and 5\farcs87 for knot M',
by their proper motion. Their kinematic ages, 590 and 525 yr, are equal
within the measurement errors, giving an average kinematic age of
550 yr for this pair of knots. This kinematic age should be considered as an
upper limit to the actual age of the knots, because these knots were most
likely formed at large distances from the central star, at the tips of conical
shocks, and because of their continuing strong deceleration by a more slowly
moving material in the bipolar lobes of He 3-1475.

\section{JET KINEMATICS WITH STIS}

    Medium-resolution STIS spectra in the vicinity of H$\alpha$ were obtained
on 15-16 June 1999. The G750M grating centered on 6581{\AA} was used with the 0\farcs1 
slit oriented along the jet axis. Spectra were obtained at nine parallel positions, 
with off-sets from the central star stepping from -0\farcs4 to 0\farcs4, so that the
5th position was centered on the star. To aid in removing cosmic ray hits, two exposures 
of about 20 min were taken at each position. In Figure 2 we show the spectrum in the
vicinity of H$\alpha$ and [N~II]. The second panel, Figure 2(W), is the WFPC2 image 
of the nebula on the same scale as the STIS spectra, while Figure 2(L) shows the
sum of slits with displacements of -0\farcs4, -0\farcs3, and -0\farcs2. Figure 2(R) 
shows displacements of 0\farcs2, 0\farcs3 and 0\farcs4, while Figure 2(C)
is from slits displaced -0\farcs1, 0\farcs0, and 0\farcs1. The vertical lines on the
WFPC2 image mark the edges of these three slit groups. The logarithm of the intensity
is shown to reveal the fainter features. 

\begin{figure}
\epsscale{0.8}    
\plotone{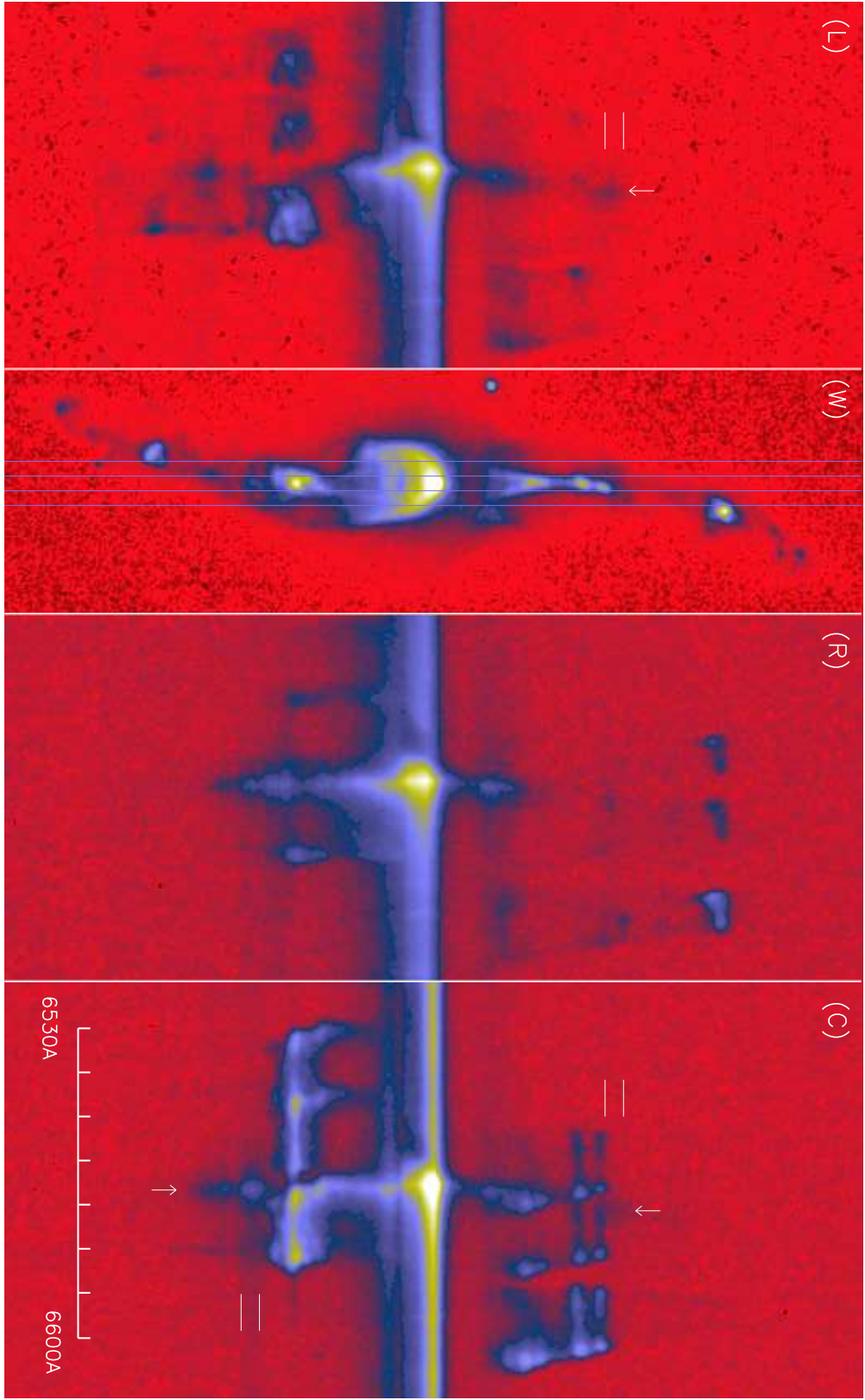}
\caption{STIS spectra of He 3-1475 in the vicinity of H$\alpha$. Panel (W)
is the WFPC2 image through the F658N [N\,II] $\lambda$6584 filter on the
same spatial scale. Panels (L),(R) and (C) are extracted from the left, right
and center regions, respectively, bounded by the vertical lines in panel (W). }
\label{fig::stis}
\end{figure}

By binning STIS spectra, we did not utilize the full spatial resolution of the 
STIS dataset in the direction perpendicular to the jets. This is not required in 
the present work, where we are mostly concerned with the global properties of 
He 3-1475, and the degradation of the spatial resolution from 0\farcs1 to 0\farcs3
is offset by an improved counting statistics. Detailed studies of the 
spatio-kinematic structure of jets, which utilize the full HST spatial resolution, 
will be presented by us in future publications.

\subsection{The Stellar Radial Velocity}

From the central spectrum which includes the star, we have measured the radial  
velocities of the stellar Fe II $\lambda$6517.88, $\lambda$6434.46, 
$\lambda$6458.17 and $\lambda$6371.22 lines, as
as well as [O~I] $\lambda$6302.045 and H$\alpha$. From the first five, we obtain 
consistent values of 37.6$\pm$1.2 km~s$^{-1}$. The central peak of H$\alpha$ yields 
54.4 km~s$^{-1}$, but this broad, asymmetric line is presumably formed in the optically
thick stellar wind, with absorption eating into the blue side of the profile. The
velocity of the Fe~II lines is close to the centroid of the 1667 MHz OH maser emission,
+45 km~s$^{-1}$ (Bobrowsky et al. 1995). Riera et al. (1995) deduced a higher systemic
velocity of +71 km~s$^{-1}$ from the centroid of the jet emission. But since the jet
velocities are $\sim$1000 km~s$^{-1}$, a small fractional asymmetry in the jet outflow 
will degrade the accuracy of this approach. We adopt +40 km~s$^{-1}$ as the systemic 
radial velocity.

\subsection{The Kinematics of the Gas}

   The first step in untangling the complex structure seen in Figure 2 is the realization 
that there are two sorts of features present: emission (presumably shock-excited) from
the jets and knots, and reflected light from the central star. The emission is 
characterized by the presence of the $\lambda$6548 and $\lambda$6583 lines of [N~II]
flanking H$\alpha~\lambda$6563. The reflected light features, on the other hand, are
characterized by H$\alpha$ with a faint continuum, but without [N~II], since the central 
star has a strong H$\alpha$ emission line (as well as many weaker emission lines of Fe~II, 
etc.).

   The most obvious features are the complex emission triplet seen in Figure 2(C).    
The upper section, above the central star, shows the red-shifted emission of the
receding jet. The [N~II] emission shows its maximum velocity of $\sim$945(905) km~s$^{-1}$
where the conical shocks emerge from the shadow of the torus; the velocity then
declines as we move to the cone tip and then to the pair of knots, dropping to $\sim$
785(745) km~s$^{-1}$ at the end. (The velocities in parenthesis are relative to the
systemic velocity of 40 km~s$^{-1}$.) All along the jet, however, we see a smear of emission
from the maximum red-shift toward much lower velocities. This is especially pronounced
beyond the cone tip at the knots. Clearly, all along the flow, material is being
decelerated from velocities of $\sim$850(810) km~s$^{-1}$ down to $\sim$200(160) km~s$^{-1}$
within a very small spatial volume. Careful inspection shows that the ``deceleration 
streaks'' are not exactly horizontal: the lower velocity gas is a bit nearer the star.
This is the head-tail structure referred to earlier. The effect is small: even where it
is most noticeable (in the first knot beyond the cone tip of the receding jet) it only 
amounts to $\sim 0\farcs05$ (i.e., about one STIS pixel).

   Below the star, we see a corresponding pattern in the blue-shifted emission from
the approaching jet. Here, there is some confusion because the reflected H$\alpha$
near the star blends with the blue-shifted [N~II] $\lambda$6583 line, but the [N~II]
$\lambda$6548 and H$\alpha$ emission reveals what is happening. From where the emission
first appears, it seems to increase in velocity slightly, reaching a maximum blue-shift
of $\sim$--910(-950) km~s$^{-1}$ just before the end of the conical structure. There is a
small drop near the cone tip, but when the bright knot is encountered, a great smear 
of emission stretches from $\sim$--795(-830) km~s$^{-1}$ down to $\sim$--125(-165) km~s$^{-1}$.
Clearly, the knots are regions of violent deceleration. Closer examination, however,
shows that the lower velocity emission is (faintly) present all along the jet. This
suggests some sort of continuous turbulent interaction along a jet channel within a
slower ambient gas.

  Beyond the knot, we see an isolated line (marked by the arrow) -- H$\alpha$ without
any corresponding [N~II] emission, which indicates that this is starlight reflected
from neutral, dusty gas. It is slightly red-shifted, an important point to which we
will return in the next section.

  Turning to Figure 2(R), we see that these slits miss the receding
jet but do just graze knot M, where we see the emission triplet of H$\alpha$
and [N~II]. The maximum red-shift out here is lower, $\sim$625(585) km~s$^{-1}$. We see
the same smear of emission down to $\sim$280(240) km~s$^{-1}$. There is an interesting
pattern here: if we take the maximum red-shift to represent the primary jet velocity,
and the lowest red-shift to be related to the velocity of the ambient gas, then as
we move away from the star, the jet velocity declines, while the velocity of the 
ambient gas increases. This inference is supported by the velocity pattern seen in
reflection discussed in the next section. 

  Finally, Figure 2(L) shows the 3 slits which just touch the bright approaching
knot. Unfortunately, the blue-shifted knot M' is beyond the last slit, though there
is a bit of emission seen.

\section{THE JET INCLINATION}

 Important information can be obtained by examining the Doppler shifts of the reflected
stellar H$\alpha$ line. Consider the approaching jet. The reflected H$\alpha$ is marked
by the lower arrow in Figure 2(C). We have extracted the spectrum 
from the brightest part, those rows between the horizontal lines, and this data is plotted
as the triangles in Figure 3. The profile of the stellar H$\alpha$ emission line was taken
from the slit passing through the star; using this stellar profile as a template with an  
adjustable amplitude and wavelength, we employed a Levenberg-Marquardt least-squares
algorithm to produce the fit plotted as the solid line in Figure 3. Note that the reflected
light has the same asymmetry as the stellar line. 
(A ``bump'' at about 6580 \AA, which is not seen in the stellar spectrum, 
is the low-velocity end of the blueshifted 6583 \AA\ [N II] emission line 
from the jet, and not the reflected stellar light.)
The result of this fit shows that the
reflected light is red-shifted by $V_{near}$ = 32.4 km~s$^{-1}$ (relative to the stellar feature).
This velocity represents the difference between a red-shift of $V$ due to the motion of the dust
directly away from the star, and a blue-shift of $V \cos(i)$ due to the component of
that motion directed towards us. Here, $i$ is the inclination of the jet axis to the
line of sight. 
  On the other hand, light reflected from the receding jet will show a larger red-shift:
$V + V \cos(i)$. In Figure 2(C) and (L), the upper arrow marks the faint reflected H$\alpha$ line 
in the region just beyond the knot emission lines; the horizontal marks show the region we
are discussing. Figure 2(L) shows the reflected line in the receding jet most clearly, since
slits 1, 2 and 3 just miss the emission from the shock cone, resulting in less confusion. 
We have extracted the spectrum from the middle slit of the three which are combined in panel
2(L), the one with the -0\farcs3 offset, shown in Figure 4.         
The red-shift is clear in comparison with the unshifted stellar H$\alpha$
(the dashed line). The least squares fit (the solid line) found a red-shift $V_{far}$ =
248 km~s$^{-1}$ relative to the stellar line. $V_{far}$ and $V_{near}$ were extracted at
the same angular distance from the star, 3\farcs9. Under the assumption that the velocity
field of the dusty neutral gas is symmetric about the central star, so that the velocity
$V$ is the same for the far and near jets, we obtain a simple result 
for the inclination: $\cos(i) = (V_{far} - V_{near})/(V_{far} + V_{near})$. We thus 
determine the inclination to be 40\degr. The true velocity at this distance from the star
is then $V = (V_{far} + V_{near})/2$ = 140 km~s$^{-1}$.

  Though it becomes fainter, the reflected H$\alpha$ extends further from the star. It 
is clear that the red-shift on both sides increases with distance from the star. Fitting
the stellar profile to the reflected line at an angular distance of 5\farcs3, we find
$V_{near}$ = 54 km~s$^{-1}$ and $V_{far}$ = 353 km~s$^{-1}$. This leads to nearly the 
same inclination, $i$ = 43\degr, and $V$ = 200 km~s$^{-1}$. While we should give less
weight to these values, the consistency of the inclination angle is reassuring. It also  
appears that the increase in $V$ with distance from the star is approximately linear.

\begin{figure}
\epsscale{0.70}    
\plotone{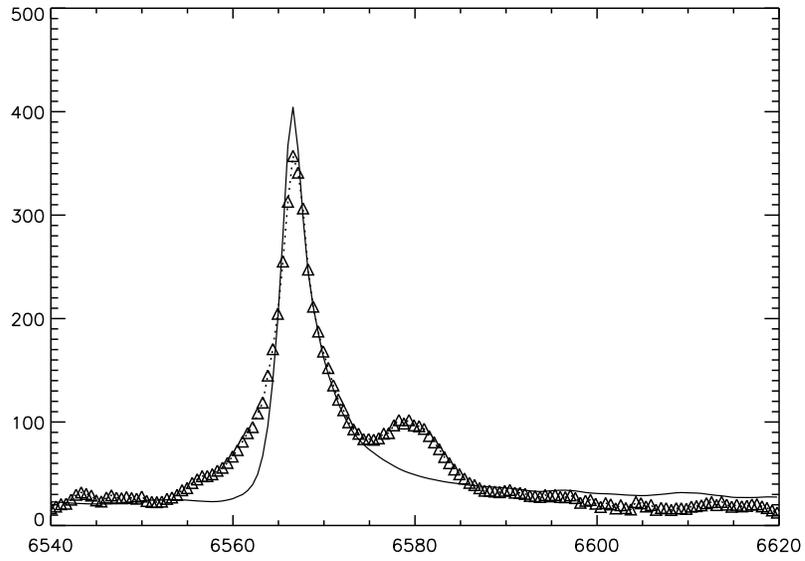}
\caption{The stellar H$\alpha$ (solid line) fit to the reflected H$\alpha$
line (triangles) of the approaching jet. }
\label{fig::ref1}
\end{figure}

\begin{figure}
\epsscale{0.70}    
\plotone{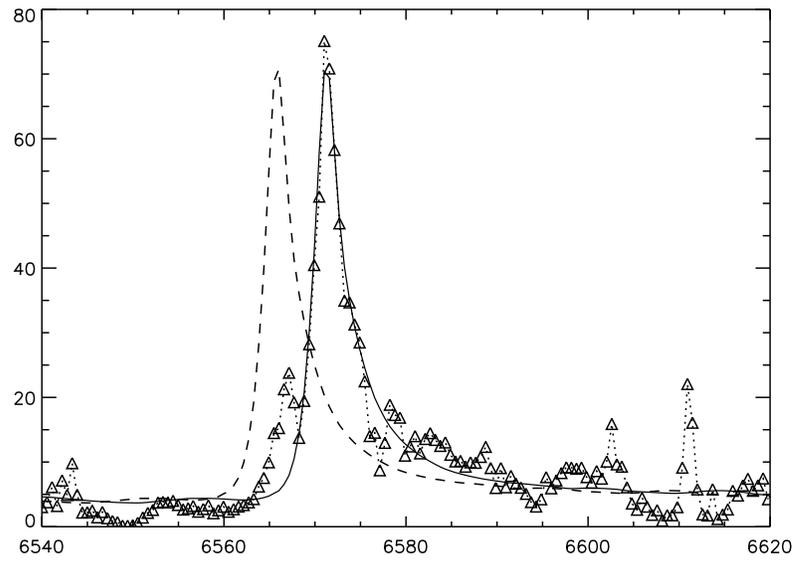}
\caption{The stellar H$\alpha$ (solid line) fit to the reflected H$\alpha$
line (triangles) of the receding jet. The dashed line is the unshifted stellar
profile.}
\label{fig::ref2}
\end{figure}

We can also obtain a limit on the inclination angle without relying on the assumed
symmetry of the near and far jets. While the ionized gas which produces the emission
lines spans a large range in velocity, the mechanism is one in which the fast-moving
jet is decelerated by its interaction with the slower neutral gas. Thus the velocity
of even the slowest ionized gas will be equal to or larger than the velocity of the neutral
gas. (It need not be an equality since it is possible that the gas stops emitting [N~II]
line radiation when its velocity difference with the neutral gas drops below a certain
value.) Thus we write $V_{ion}/V_{neut} = R \geq 1$, where $V_{ion}$ and $V_{neut}$ are
the velocities, along the jet axis, of the slowest optically emitting gas and of the neutral 
gas, respectively. Then we find that

$V_{neut} = V_{near} - V_{emiss}/R ~~~~~~~~\hbox{and}~~~~~~~\cos(i)~=~\left[1~+~R V_{near}/
       (-V_{emiss})\right]^{-1}$

\noindent where $V_{emiss} = V_{emiss}' - V_{star}$, the observed Doppler shift 
of the slowest emitting gas relative to the star. By setting $R = 1$ we find the
minimum inclination angle. From the the [N~II] emission just beyond the bright knot
in the approaching jet, we find $V_{emiss}' = -150$ km~s$^{-1}$, so $V_{emiss} = -187$
km~s$^{-1}$. With $V_{near} = 32$ km~s$^{-1}$, the inclination is 
$i =$ 31\fdg5, and $V_{neut} = V_{ion} = 219$ km~s$^{-1}$. 

Note, however, that an inclination of 31\fdg5 implies a strong asymmetry in the 
jets: For the receding jet we require $V_{neut} = 134$ km~s$^{-1}$
to fit the observed $V_{far}$ of the reflected H$\alpha$ line, which is 40\%\
less than the velocity at the corresponding distance on the approaching side. 
Such a large difference seems in conflict with the overall symmetry we see in
the emission line velocities of the near and far jets.
On the other hand, with $R = 1.8$ we recover the symmetric solution, $i =$ 
40\degr, with $V_{neut} = 137$ km~s$^{-1}$ and $V_{ion} = 244$ km~s$^{-1}$. We prefer
this solution. If this interpretation is correct, then it follows that the jet
flow stops emitting strongly when its velocity difference with the neutral gas
drops below $\sim$100 km~s$^{-1}$.

\section{THE DISTANCE TO HE 3-1475}

We may now find the distance to He 3-1475 by comparing the measured proper 
motions of knots M and M' with their tangential velocities. Our STIS
spectra just graze the red-shifted knot M (Fig. 2(R)), where we see emission line 
velocities between 240 and 585 km~s$^{-1}$. Because the spectrum offset 0\farcs4 
shows higher velocities than that offset 0\farcs3, and since no STIS spectra
pass through the knot center, we likely underestimate the true knot velocities.
In addition, we have no STIS spectra of the blue-shifted knot M'. We therefore use
the ground-based radial velocities measured for knots M and M', $v_r$ of
$\pm 500$ km s$^{-1}$ (Riera et al. 1995; Bobrowsky et al. 1995), 
to circumvent this
problem. The implied tangential velocities $v_T$ are $v_r \cot 40$\degr$ = 420$ 
km s$^{-1}$. Since we determined in \S~2 that the average radial proper motion
of knots M and M' is 10.5 miliarcsec yr$^{-1}$, we arrive at a distance of
8.3 kpc. The galactic coordinates of He 3-1475 are $l = 9$\fdg4 and
$b = +5$\fdg8, which in combination with the distance of 8.3 kpc implies
that this star is located in the Galactic Bulge, 800 pc above the Galactic
Plane. 

We should note that we are making the assumption that our measured proper       
motions of the knot images correspond to the physical velocity of the emitting
gas determined spectroscopically. That is, the knots are not just some sort of
wave propagating along the jet. This seems plausible because the morphology of
the middle knots is well defined and does not show any change between the epochs 
of observation. It would be useful, however, to observe the knots over a longer 
time interval. Detailed hydrodynamic modeling might also provide insights.

The velocity of the (neutral) bipolar outflow, as determined from the scattered stellar 
H$\alpha$ line, increases approximately linearly with the distance from the 
central star. We can then determine the kinematic age of this outflow from
any knot within the bipolar outflow by dividing its projected distance from 
the central star by its tangential velocity. For example, the 140 km s$^{-1}$ 
radial velocity of the brightest knot seen in the scattered light in 
the blue-shifted lobe (at distance of 3\farcs9 from the star) translates
into tangential velocity of 90 km s$^{-1}$. We then obtain an age of 
$1,700$ yr at distance of 8.3 kpc. This age is several times longer than 
the age of the knots within the jet. We obtain a similar (1,300 yr) 
age for the material 
seen in the OH maser emission, assuming that the molecular outflow with 
a radial expansion velocity of 25 km s$^{-1}$ and a spatial extent of
2\arcsec\ (Bobrowsky et al. 1995) is perpendicular to the jet. The bipolar
outflow and the OH outflow apparently originated at the same time, perhaps 
at the time when the central star ejected the torus seen in the HST images,
while the jet activity is more recent. This jet activity must be continuing
now, as indicated by the short (120 yr) stellar wind travel time 
from the star to the cone tips.

\section{DISCUSSION}

Our distance determination places He 3-1475 in the Galactic Bulge, at
a much larger distance than previously anticipated. At this distance its
infrared luminosity is equal to 25,000 $L_\odot$ (Riera et al. 1995), which 
might be less that the stellar luminosity if the dust optical depth
is low along the bipolar outflow so that a substantial fraction of stellar
luminosity escapes along the poles. He 3-1475 is apparently a high-luminosity
Galactic Bulge star at a height of 800 pc above the Galactic plane. Its 
location suggests 
that He 3-1475 is an old Population II star, but an old, low-mass post-AGB 
star 
cannot be so luminous -- a more massive star is required. Luminous,
more massive AGB stars are indeed found in old stellar populations 
(e. g., in the globular cluster NGC 6553 -- Guarnieri et al. 1997), although 
their luminosities may be several times lower than that of He 3-1475. It is
also feasible that He 3-1475 is a member of a trace intermediate-age population
of Galactic Bulge.

Irrespective of its age and origin, He 3-1475 appears to be an evolved star 
which lost a substantial amount of mass in a slow asymmetric wind at least
1,500 yr ago, possibly in a superwind at the AGB tip or during a binary 
interaction, and which is now seen in the optical as a dense torus. 
The presence of broad P-Cygni profiles in the stellar 
spectrum shows that the transition from the slow to the fast wind has
occurred since then. The B[e] spectral classification of the central star
implies that this fast wind is asymmetric, according to a disk wind model
for B[e] stars by Zickgraf et al. (1985). This wind is perhaps driven by 
stellar UV radiation from a single fast rotating B type star (e. g., 
Pelupessy, Lamers, \& Vink 2000), or more generally from 
a star + a disk system (Oudmaijer et al. 1998). This asymmetric wind 
is then collimated into 1200 km s$^{-1}$ jets during its interaction with the more 
slowly moving material ejected earlier, maybe purely by hydrodynamical 
means (Borkowski et al. 1997). The time variability often seen in many B[e] 
stars suggests that a ``knotty'' jet morphology seen in He 3-1475 could
be caused by temporal variations in the stellar wind properties. 

Our STIS observations show how this high-velocity, time-dependent jet is 
decelerated by the circumstellar medium on both large and small scales. We
see the large scale deceleration in the decrease of the maximum velocity
as we move down the jet and out to the knots. But we also see at all points
along the flow, and especially at the knots, local ($\sim 0\farcs05 = 6 
\times 10^{15}$ cm) decelerations of over 500 km s$^{-1}$. Consider 
that a plane-parallel shock of 500 km s$^{-1}$ would produce gas with a
temperature of $3 \times 10^6$ K, while we see a spectrum which seems more
appropriate for a cooling shock of far lower velocity, about 100 km s$^{-1}$
(Riera et al. 1995). 
Clearly, there is much hydrodynamical structure which is still unresolved
by HST.

It is still a mystery why B[e] star+disc systems are present in evolved 
post-AGB stars such as He 3-1475. There are a dozen such systems in compact
PNe or proto-PNe (Lamers et al. 1998), albeit generally of lower luminosity 
than He 3-1475. The best known of these systems is the ``Butterfly'' PN M2-9,
a low luminosity (550 L$_\odot$) object at a distance of 650 pc 
(Schwarz et al. 1997). M2-9 is clearly very much different than He 3-1475,
suggesting that compact PN B[e] central stars form a heterogeneous class,
just like B[e] stars as a whole (Lamers et al. 1998). M2-9 shows a complex
kinematics, with young ($< 10$ yr) micro-jets close to the star 
with velocities up to 195 km s$^{-1}$ and much older (1,300 yr) bipolar 
outflows with velocities up to 140 km s$^{-1}$ (Solf 2000). The 
1200 yr old outermost reflection knots are expanding with velocity of
165 km s$^{-1}$ (Schwarz et al. 1997). 
Doyle et al. (2000) find evidence for $\sim 1000$ km s$^{-1}$ outflows.
High velocity (500 km s$^{-1}$) outflows were also found by 
Redman et al. (2000) in another PN with a B[e] central star, Mz3, while
spectacular jets were recently imaged by the HST in yet another PN on 
Lamers' et al. PN and PPN list, He 2-90 (Sahai \& Nyman 2000). 
Both M2-9 and He 2-90 are suspected to be interacting binaries.
An interacting binary progenitor hypothesis also seems attractive for
He 3-1475, because its very high luminosity excludes a typical low-mass
AGB star in the Galactic Bulge. A mass transfer binary and/or a merger event
provides a reasonable solution to the presence of a more massive star in the
Galactic Bulge.

In this paper, we concentrated our attention on the large-scale jet kinematics
and on the global properties of He 3-1475. The next step in analysis of the
extensive STIS dataset should involve detailed studies of individual knots
and of the jet collimation regions. Such studies are essential for 
providing answers to questions about the mechanism of the jet collimation,
about the nature of the observed shock emission, and about the temporal and
spatial variability of the flow. For example, a simple (but perhaps naive)
interpretation would be to attribute the observed shock emission to shocks
driven into the jets, such as encountered in working surfaces of continuous
jets. Because of the high jet velocities, the bow shocks ahead of the 
knots should in principle be nonradiative,
X-ray emitting shocks, with little or no optical emission. The presence of
distinct knots symmetrically distributed with respect to the star suggests 
that the stellar outflow varies on timescales from tens to hundreds 
of years, leading to noncontinuous, episodic, and wobbling jets. It is also
not certain whether the jets can be collimated by purely 
hydrodynamical processes, or whether magnetic fields must be involved. 
We hope that the continuing analysis of the HST He 3-1475 dataset will provide
answers to some of these questions in the near future.

\acknowledgements We would like to thank Daniel Proga for stimulating
discussions about the nature of B[e] stars. We also thank the referee,
Romano Corradi, for useful comments which led to substantial improvements in 
the presentation of our results.
Support for this work was provided by NASA through grants GO-07285.01-96A 
and GO-07285.02-96A from 
the Space Telescope Science Institute, which is operated by the Association of
Universities, Inc., under NASA contract NAS5-26555.

\end{document}